\documentclass[aip,rsi, amsmath,amssymb, reprint, fontsize=12pt]{revtex4-1}
\pdfoutput=1
\usepackage{graphicx}
\usepackage{dcolumn}
\usepackage{bm}

\usepackage{pgfplots}
\usepackage{pgfplotstable}
\usepgfplotslibrary{external} 
\usepgfplotslibrary{fillbetween}
\usepgfplotslibrary{colorbrewer}
\pgfplotsset{compat=1.6}
\usepackage{lineno}

\usetikzlibrary{external}
\usepackage{subfigure}
\usepackage{cleveref}

\pgfmathsetmacro{\Tfreeze}{364}
\pgfmathsetmacro{\Tjump}{442.335666891}
\pgfmathsetmacro{\tfreeze}{0.483233608052}
\pgfmathsetmacro{\tjump}{0.535762818744}
\pgfmathsetmacro{\Tc}{537}

\definecolor{android_blue}{RGB}{51,181,229}
\definecolor{android_dark_blue}{RGB}{0,153,204}
\definecolor{android_pink}{RGB}{170,102,204}
\definecolor{android_purple}{RGB}{156,39,176}
\definecolor{android_dark_pink}{RGB}{153,51,204}
\definecolor{android_green}{RGB}{153,204,0}
\definecolor{android_dark_green}{RGB}{102,153,0}
\definecolor{android_orange}{RGB}{255,152,0}
\definecolor{android_dark_orange}{RGB}{255,152,0}
\definecolor{android_red}{RGB}{255,68,68}
\definecolor{android_dark_red}{RGB}{204,0,0}
\definecolor{android_pink}{RGB}{156,39,176}
\definecolor{android_grey}{RGB}{158,158,158}

\definecolor{olivia_red}{RGB}{215,25,28}
\definecolor{olivia_orange}{RGB}{253,174,97}
\definecolor{olivia_yellow}{RGB}{255,255,191}
\definecolor{olivia_lightblue}{RGB}{171,217,233}
\definecolor{olivia_blue}{RGB}{44,123,182}

\pgfplotsset{grid style={dashed,grey,opacity=0.5}}

\pgfplotscreateplotcyclelist{peak_temp}{
{color=android_dark_blue,line width=1.0pt,mark=x,mark size=2pt,mark options={line width=1.0pt},line join=round},
{color=android_red,line width=1.0pt,mark=o,mark size=2pt,mark options={line width=1.0pt},line join=round},
{color=android_dark_green,line width=0.5pt,mark=|,mark size=2pt,mark options={line width=0.75pt},line join=round},
{color=black,line width=0.75pt,mark size=2pt,mark=triangle,mark options={line width=0.75pt},line join=round},
{color=android_blue,line width=0.5pt,mark=square,mark size=2pt,mark options={line width=0.75pt},line join=round},
	  {color=android_pink,line width=0.5pt,mark=diamond,mark size=2pt,mark options={line width=0.75pt},line join=round},
	  {color=android_orange,line width=0.5pt,mark size=2pt,mark options={line width=0.75pt},line join=round}
	  }

\begin{document}
\pgfplotsset{colormap/RdBu-9}

\title{The superior role of the Gilbert damping on the signal-to-noise ratio in heat-assisted magnetic recording }

\author{O. Muthsam}
 \email{olivia.muthsam@univie.ac.at}
\author{F. Slanovc}
\author{C. Vogler}
\author{D. Suess}
\affiliation{ 
University of Vienna, Physics of Functional Materials, Boltzmanngasse 5, 1090 Vienna, Austria
}%

\date{\today}
             
\begin{abstract}
In magnetic recording the signal-to-noise ratio (SNR) is a good indicator for the quality of written bits. However, a priori it is not clear which parameters have the strongest influence on the SNR. In this work, we investigate the role of the Gilbert damping on the SNR. Grains consisting of FePt like hard magnetic material with two different grain sizes $d=5\,$nm and $d=7\,$nm are considered and simulations of heat-assisted magnetic recording (HAMR) are performed with the atomistic simulation program VAMPIRE. 
The simulations display that the SNR saturates for damping constants larger or equal than 0.1. Additionally, we can show that the Gilbert damping together with the bit length have a major effect on the SNR whereas other write head and material parameters only have a minor relevance on the SNR.
\end{abstract}
\maketitle
\section{Introduction}
The next generation recording technology to increase the areal storage density of hard drives beyond 1.5\,Tb/in$^2$ is heat-assisted magnetic recording (HAMR) \cite{burns,thermomagnetic,kryder,rottmayer,kobayashi,mee}. Higher areal storage densities (ADs) require smaller recording grains. These grains need to have high anisotropy to be thermally stable. HAMR uses a heat pulse to locally enhance the temperature of the high anisotropy recording medium beyond the Curie temperature. Due to the heating, the coercivity of the grain drops and it can be written with the available head fields. After the grain is written, the medium is cooled and the information is safely stored. A good indicator for the quality of the written bits is the so-called signal-to-noise ratio (SNR) which gives the power of the signal over the power of the noise \cite{zhu2013understanding}. To achieve high areal storage densities, recording materials that show good magnetic properties even at small grain sizes and thus yield high SNR values are needed. However, a priori it is not clear which parameters have the strongest influence on the SNR.\\
In this work, we investigate the effect of a varying damping constant on the SNR. HAMR simulations with the atomistic simulation program VAMPIRE \cite{evans} are performed for cylindrical recording grains with two different diameters $d=5\,$nm and $d=7\,$nm and a height $h=8\,$nm. The material parameters of FePt like hard magnetic recording media according to the Advanced Storage Technology Consortium (ASTC) \cite{ASTC} are used. Damping constants between $\alpha=0.01$ and $\alpha=0.5$ are considered.
Additionally, we present an equation to include the influence of the bit length to the SNR. With this we can explain a SNR decrease of about 8.25\,dB for 5-\,nm grains, which results when changing the material and writing parameters in the HAMR simulations from those used in former simulations \cite{areal,fundamental,noisehamr} to those according to the Advanced Storage Technology Consortium \cite{ASTC}, with the damping constant and the bit length only.\\ 
The structure of this paper is as follows: In Section II, the HAMR model is introduced and it is explained how the SNR is determined. In Section III, the results are presented and in Section IV they are discussed.
\section{HAMR Model}
Cylindrical recording grains with height $h=8\,$nm and diameters $d=5\,$nm  and $d=7\,$nm are considered. One grain can be interpreted as one grain of a state-of-the-art granular recording medium. A simple cubic crystal structure is used. The exchange interaction $J_{ij}$ and the effective lattice parameter $a$ are adjusted so that the simulations lead to the experimentally obtained saturation magnetization and Curie temperature \cite{mryasov2005temperature,hovorka2012curie}. In the simulations, only nearest neighbor exchange interactions between the atoms are included. A continuous laser pulse with Gaussian shape and the full width at half maximum (FWHM) of 60\,nm is assumed in the simulations. The temperature profile of the heat pulse is given by
\begin{align}
T(x,y,t)= (T_{\mathrm{write}}-T_{\mathrm{min}})e^{-\frac{x^2+y^2}{2\sigma^2}} + T_{\mathrm{min}} \\
= T_{\mathrm{peak}}(y)\cdot e^{-\frac{x^2}{2\sigma^2}} + T_{\mathrm{min}}
\label{pulse}
\end{align}
with
\begin{align}
\sigma=\frac{\mathrm{FWHM}}{\sqrt{8\ln(2)}}
\end{align}
and
\begin{align}
T_{\mathrm{peak}}(y)=(T_{\mathrm{write}}-T_{\mathrm{min}})e^{-\frac{y^2}{2\sigma^2}}.
\label{equation}
\end{align}
$v=15\,$m/s is the speed of the write head. $x$ and $y$ label the down-track and the off-track position of the grain, respectively. In our simulations both the down-track position $x$ and the off-track position $y$ are variable. The ambient and thus minimum temperature of all simulations is $T_{\mathrm{min}}=300\,$K. The applied field is modeled as a trapezoidal field with a field duration of 0.57\,ns and a field rise and decay time of 0.1\,ns, resulting in a bit length of 10.2\,nm. The field strength is assumed to be $+0.8$\,T and $-0.8$\,T in $z$-direction. Initially, the magnetization of each grain points in $+z$-direction. The trapezoidal field tries to switch the magnetization of the grain from $+z$-direction to $-z$-direction. At the end of every simulation, it is evaluated if the bit has switched or not.\\
The material and write head parameters according to the Advanced Storage Technology Consortium \cite{ASTC} are shown \Cref{SNRtablematerialien}. 
\begin{center}
\begin{table*}
\centering
\begin{tabular}{|>{\centering}m{2.cm}|>{\centering}m{2.cm}|>{\centering}m{3.cm}|>{\centering}m{2.0cm}|>{\centering}m{1.5cm}|c|>{\centering}m{2.0cm}|c|}
\hline
Curie temp. $T_{\mathrm{\textbf{C}}}$ [K] & Damping $\alpha$& Uniaxial anisotropy $k_{u}$ [J/link]& $J_{ij}$ [J/link]& $\mu_{\mathrm{s}}$  [$\mu_{\mathrm{B}}$]  & $v$\,[m/s]&  field duration (fd)\,[ns]&  FWHM\, [nm]\\
    \hline
	693.5 & 0.02&$9.124\times10^{-23}$  & $6.72 \times 10^{-21}$ &1.6&15& 0.57&60\\
    \hline
 \end{tabular}
\caption{Material and write head parameters of a FePt like hard magnetic granular recording medium accoring to the Advanced Storage Technology Consortium. }
\label{SNRtablematerialien}
\end{table*}
\end{center}
\subsection{Determination of SNR \label{section:SNR}}
To calculate the signal-to-noise ratio, the read-back signal of a written bit pattern has to be determined. To write the bit pattern and get the read-back signal from it, the following procedure is used. First, a switching probability phase diagram is needed for the writing process of the bit pattern. Since it is very time consuming to compute a switching probability phase diagram with atomistic or micromagnetic simulations, an analytical model developed by Slanovc \textit{et al} \cite{slanovc} is used in this work. The model uses eight input parameters (the maximum switching probability $P_{\mathrm{max}}$, the down-track jitter $\sigma_{\mathrm{down}},$ the off-track jitter $\sigma_{\mathrm{off}},$ the transition curvature $c$, the bit length $b$, the half maximum temperature $F_{50}$, the position $p_2$ of the phase diagram in $T_{\mathrm{peak}}$ direction and the position $p_3$ of the phase diagram in down-track direction) to determine a switching probability phase diagram. Slanovc \textit{et al} showed that the maximum switching probability $P_{\mathrm{max}}$ and the down-track jitter $\sigma_{\mathrm{down}}$ are the input parameters with the strongest influence on the SNR. Note, that the bit length $b$ also has a strong influence on the SNR. In the further course of this work, an equation to include the bit length to the SNR calculations is shown. Thus, the bit length can be assumed constant during the SNR determination. The transition curvature $c$ did not show strong influence on the SNR for the used reader model and the off-track jitter $\sigma_{\mathrm{off}}$ is neglectable since the reader width is with $30.13\,$nm smaller than the track width with $44.34\,$nm and thus does not sense the off-track jitter. $p_2$ and $p_3$ only shift the bit pattern and can thus be fixed for comparability. For this reason, it is reasonable to fix the input parameters, except for the maximum switching probability $P_{\mathrm{max}}$ and the down-track jitter $\sigma_{\mathrm{down}}$. The fixed input parameters are determined by a least square fit from a switching probability phase diagram  computed with a coarse-grained Landau-Lifshitz-Bloch (LLB) model \cite{vogler_landau-lifshitz-bloch_2014} for pure hard magnetic grains with material parameters given in \Cref{SNRtablematerialien}. The fitting parameters are summarized in \Cref{tab::fit} for grain diameters $d=5\,$nm and $d=7\,$nm.\\
Further, it is necessary to compute the down-track jitter $\sigma_{\mathrm{down}}$ and the maximal switching probability $P_{\mathrm{max}}$ for the considered set of material and write head parameters, see \Cref{SNRtablematerialien}. In the simulations, the switching probability of a recording grain at various down-track positions $x$ at a peak temperature $T_{\mathrm{peak}} = T_{\mathrm{c}} + 60\,$K is calculated with the atomistic simulations program VAMPIRE \cite{evans}, yielding a down-track probability function $P(x)$.
To get the down-track jitter and the maximum switching probability, the switching probability curve is fitted with a Gaussian cumulative distribution function 
\begin{align}
\Phi_{\mu,\sigma^2}=\frac{1}{2} (1 + \mathrm{erf}(\frac{x-\mu}{\sqrt{2\sigma^2}}))\cdot P_{\mathrm{max}}
\label{distribution}
\end{align}
with
\begin{align}
\mathrm{erf}(x)=\frac{2}{\sqrt{\pi}} \int_0^x e^{-\tau^2} d\tau,
\label{error}
\end{align}
where the mean value $\mu$, the standard deviation $\sigma$ and the mean maximum switching probability $P_{\mathrm{max}} \in [0,1]$ are the fitting parameters. The standard deviation $\sigma$, which determines the steepness of the transition function, is a measure for the transition jitter and thus for the achievable maximum areal grain density of a recording medium. The fitting parameter $P_{\mathrm{max}}$ is a measure for the average switching probability at the bit center.  Note, that the calculated jitter values $\sigma_{\mathrm{down}}$ only consider the down-track contribution of the write jitter. The so-called $a-$parameter is given by
\begin{align}
    a =  \sqrt{\sigma_{\mathrm{down}}^2+\sigma_{\mathrm{g}}^2}
\end{align}
where $\sigma_{\mathrm{g}}$ is a grain-size-dependent jitter contribution \cite{wang_transition_2009}. The write jitter can then be calculated by
\begin{align}
    \sigma_{\mathrm{write}} \approx a \sqrt{\frac{S}{W}}
\end{align}
where $W$ is the reader width and $S=D+B$ is the grain size, i.e. the sum of the grain diameter $D$ and the nonmagnetic boundary $B$ \cite{varvaro_ultra-high-density_2016,slanovc}. \\
For each $\sigma_{\mathrm{down}}$ and $P_{\mathrm{max}}$ combination a switching probability phase diagram is computed with the analytical model. With the resulting phase diagram, the writing process of a certain bit pattern is simulated on granular recording medium \cite{slanovc}. Here, the switching probability of the grain is set according to its position in the phase diagram. The writing process is repeated for 50 different randomly initialized granular media. Finally, the read-back signal is determined with a reader model where the reader width is 30.13\,nm and the reader resolution in down-track direction is $13.26$\,nm. The SNR can then be computed from the read-back signal with the help of a SNR calculator provided by SEAGATE \cite{hernandez_using_2017}. 
The resulting SNR value is given in dB (SNR$_{\mathrm{dB}}$). In the following, the SNR$_{\mathrm{dB}}$ is simply called SNR unless it is explicitly noted different.
\begin{table}
\begin{center}
\begin{tabular}{|c|cc|}
\hline
&  \multicolumn{2}{c|}{grain size}          \\
&  $5$ nm  & $7$ nm  \\
\hline
$\sigma_{\mathrm{off}}$ [K]  & 22.5  & 14.4 \\
$P_\text{max}$  & 0.995  & 0.997  \\
$F_{50}$ [K]  & 602  & 628  \\
$b$ [nm]  & 10.2  & 10.2  \\
$c$ [$10^{-4}$ nm/$\text{K}^2$]  & 3.88  & 4.89 \\
$p_2$ [$\text{K}$]  & 839  & 830 \\
$p_3$ [nm]  & 27.5  & 25.8 \\
\hline
\end{tabular}
\end{center}
\caption{Reference parameters that are evaluated via least square fit of the simulated phase diagrams for grain sizes $5\,$nm and $7\,$nm. Details of the parameters can be found in \cite{slanovc}.}
\label{tab::fit}
\end{table}
\section{Results}
\subsection{SNR Dependency on Damping}
\begin{figure}
\centering
\includegraphics[width=1.0\linewidth]{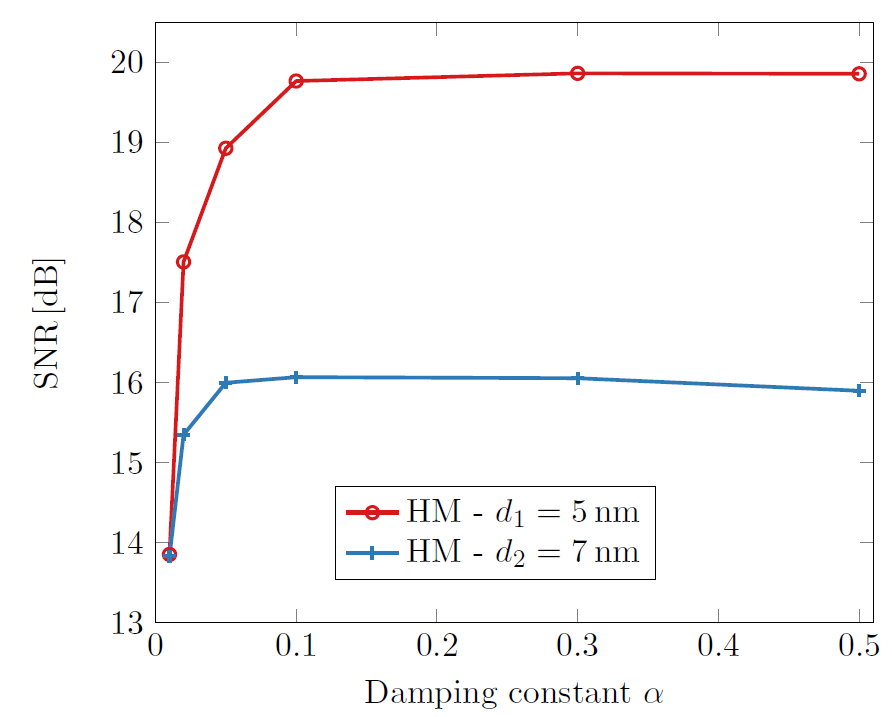}
  \caption{Resulting SNR for various damping constants $\alpha$ for grains with two different diameters $d=5\,$nm and $d=7\,$nm. }
  \label{SNRvsdamping}
\end{figure}

\begin{figure}
\centering
\includegraphics[width=1.0\linewidth]{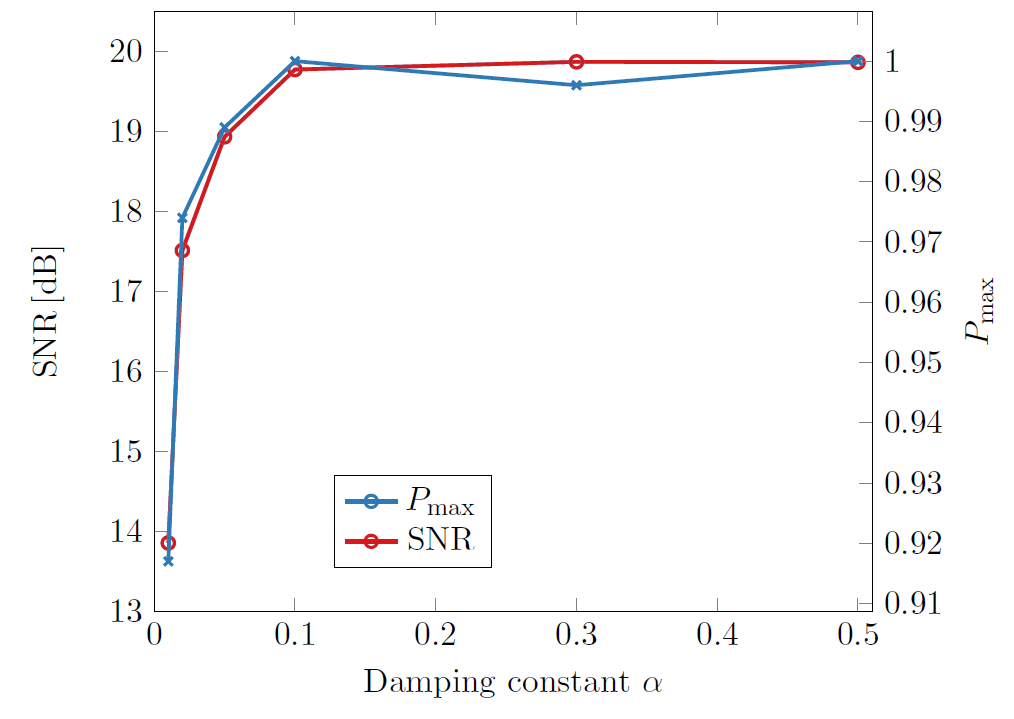}
  \caption{SNR and $P_{\mathrm{max}}$ depending on the damping constant $\alpha$ for grain size with a diameter of $5\,$nm.}
  \label{SNRPmax}
\end{figure}

\begin{figure}
\centering
\includegraphics[width=1.0\linewidth]{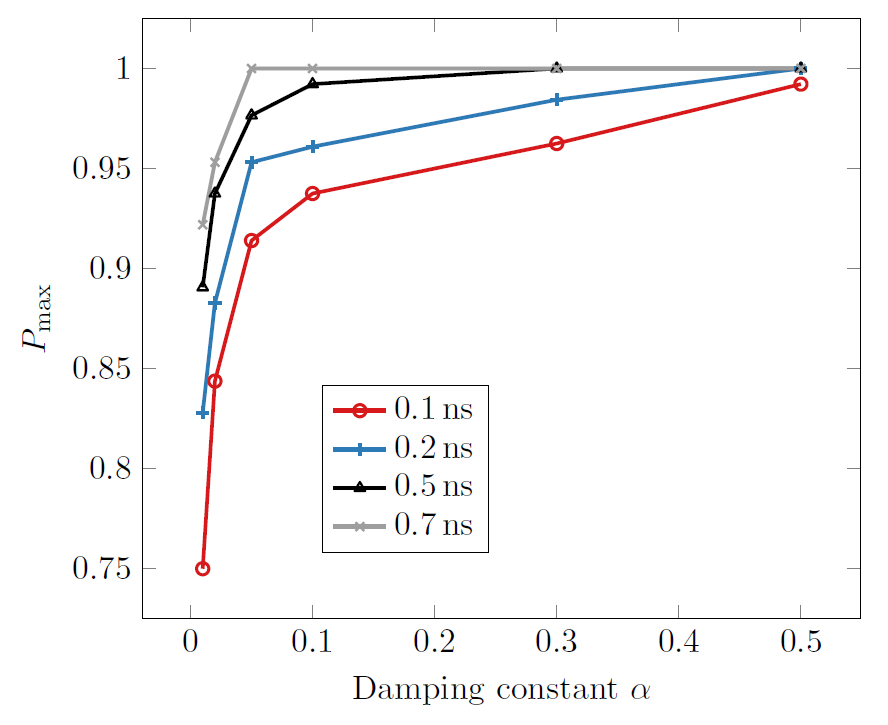}
  \caption{Maximum switching probability $P_{\mathrm{max}}$ over damping $\alpha$ for different pulse lengths of the heat pulse. A field strength of $-0.8\,$T for grains with diameter $d=5\,$nm is assumed. }
  \label{pulsewidth}
\end{figure}

\begin{figure}[h]
\centering
\includegraphics[width=1.0\linewidth]{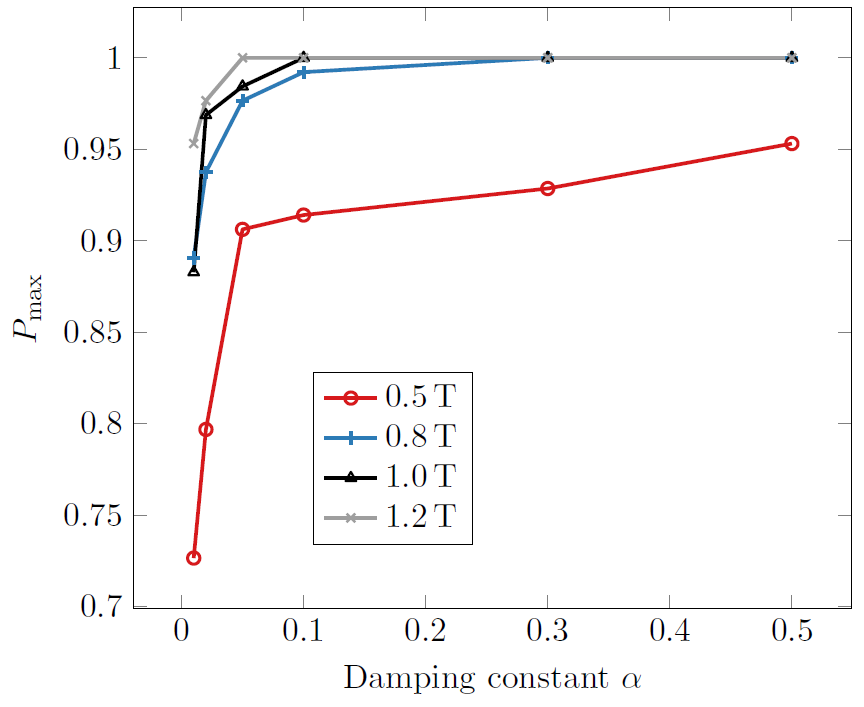}
  \caption{Maximum switching probability $P_{\mathrm{max}}$ over damping $\alpha$ for different field strengths. The durations of the heat pulse of $0.5\,$ns for grains with diameter $d=5\,$nm is assumed.  }
  \label{fieldstrengthdamping}
\end{figure}

First, the influence of the damping constant on the SNR is investigated in more detail. The damping constant is varied from $\alpha=0.01$ to $\alpha=0.5$ for two different grain sizes $d=5\,$nm and $d=7\,$nm. All other parameters are taken from \Cref{SNRtablematerialien}. The bit length in the simulations is $10.2\,$nm and the track width is $44.34\,$nm. The down-track jitter curves are computed at $T_{\mathrm{peak}}=760\,$K and fitted with \cref{distribution}. In \Cref{SNRvsdamping}, the SNR over the damping constants for both grain sizes is visible. Note that the SNR is proportional to the number of grains, meaning that the number of grains per bit has to be kept constant to determine a nearly constant SNR \cite{wood2000feasibility}. However, since the dimensions of the granular media used for the writing and reading process are fixed, less grains form one bit for $d=7\,$nm. Thus, the SNR values for the larger grain size are smaller than for the small grains.\\
The results show that changing the damping constant from $\alpha=0.01$ to $\alpha=0.02$ already increases the SNR by 3.66\,dB for 5\,nm-grains. For $d=7\,$nm, the SNR gain is 1.65\,dB. For 5\,nm-grains, damping constants $\alpha \ge 0.1$ lead to the best results with a total improvement of 6\,dB compared to $\alpha=0.01$. Surprisingly, enhancing the damping constant beyond $0.1$ does not show any further improvement, the SNR saturates. This behavior is the same for the 7\,nm-grains. However, the total betterment of the SNR is only 2.24\,dB for the larger grains. The SNR saturation results from the fact that $P_{\mathrm{max}}=1$ for $\alpha \ge 0.1$. Simultaneously, the down-track jitter $\sigma_{\mathrm{down}}$ varies only marginally for $\alpha \ge 0.1$ (see \Cref{dampingsigmaPmax}) such that it does not alter the SNR. The correlation between the SNR and the maximum switching probability $P_{\mathrm{max}}$ is shown in \Cref{SNRPmax}. It shows that the fitted SNR curve reproduces the data very well. \\
By further studying the switching dynamics of a 5\,nm-grain, one can show that the assumed pulse duration of the heat pulse and the applied field strength are crucial for the saturation of the SNR. In \Cref{pulsewidth}, it is displayed how the duration of the heat pulse influences the maximum switching probability and with it the SNR. During the duration of the heat pulse the field is considered to constantly point in $-z-$direction. The results demonstrate that $P_{\mathrm{max}}$ does not saturate for small pulse durations. If longer pulse durations $\ge 0.5\,$ns are assumed, a $P_{\mathrm{max}}$ saturation can be seen. A similar effect can be seen for a change of the field strength when the pulse duration is assumed to be $0.5\,$ns (see \Cref{fieldstrengthdamping}). For a small head field with a strength of $0.5\,$T, $P_{\mathrm{max}}$ shows no saturation whereas it does for larger head fields. From the simulations with varying duration of the heat pulse and field strength, it can also be seen that the SNR can be improved for smaller damping constants if the duration of the heat pulse is increased due to a smaller head velocity or the field strength are enhanced.
\begin{center}
\begin{table*}
\centering
\begin{tabular}{|c|c|c||c|c|c|c|}\hline
&  \multicolumn{2}{c||}{5\,nm}    &  \multicolumn{2}{c|}{7\,nm}      \\
$\alpha$ & $\sigma_{\mathrm{down}}$\,[nm] &$P_{\mathrm{max}}$\,& $\sigma_{\mathrm{down}}$\,[nm] &$P_{\mathrm{max}}$\,\\
\hline
0.01&2.0&0.917 & 1.13 & 0.955\\
 0.02&0.9475&0.974& 0.83 & 0.99\\
 0.05&0.7&0.989 & 0.549& 1.0\\
  0.1&0.688&1.0& 0.442 & 1.0\\
  0.3&0.495&1.0& 0.48 & 1.0\\
  0.5 & 0.64& 1.0& 0.636 & 1.0\\
   \hline
 \end{tabular}
\caption{Resulting down-track jitter parameters and mean maximum switching probability values for pure hard magnetic material with different damping constants $\alpha$.}
\label{dampingsigmaPmax}
\end{table*}
\end{center}
\subsection{SNR Dependency on bit length \label{section:bit length}}
The influence of the bit length on the SNR was already studied by Slanovc \textit{et al} \cite{slanovc}. In this work, the following calculation is important. For the SNR calculations a bit length $b_1=10.2$\,nm is assumed since this is the bit length resulting from the ASTC parameters. The track width in the simulations is again $44.34\,$nm. However, the bit length can change due to a variation of the write head parameters (field duration and head velocity). Therefore, the bit length for the former parameters \cite{areal,fundamental,noisehamr} is 22\,nm. To write a bit pattern with larger bit lengths ($b>12\,$nm) the simulations of new granular media are required. This is computationally very expensive. Thus, a different approach is needed to qualitatively investigate the influence of the bit length. For the SNR with SNR$_{\mathrm{dB}}=10 \log_{10}(\mathrm{SNR})$, there holds \cite{varvaro_ultra-high-density_2016}
%
\begin{align}
    \mathrm{SNR} \propto \left(\frac{b}{a}\right)^2 \left(\frac{T_{50}}{b}\right) \left(\frac{W}{S}\right)
\end{align}
with the bit length $b$ and the read-back pulse width $T_{50}$ which is proportional to the reader resolution in down-track direction. The ratio $T_{50}/b$ is called user bit density and is usually kept constant \cite{varvaro_ultra-high-density_2016}. Further, the reader width $W$ and the grain size $S$ are constant. Since the aim is to qualitatively describe the SNR for a bit length $b_2$ from SNR calculations with a bit length $b_1,$ the a-parameter $a$ is also assumed to be constant. The SNR$_{\mathrm{dB}}$ for a different bit length $b_2$ can then be calculated by

\begin{align}
    \mathrm{SNR}_{\mathrm{dB}}(b_2)-\mathrm{SNR}_{\mathrm{dB}}(b_1) \nonumber \\= 10 \log_{10}(\mathrm{SNR}(b_2))-10\log_{10}(\mathrm{SNR}(b_1)) \nonumber\\= 10 \log_{10}(b_2^2)-10 \log_{10}(b_1^2) = 20 \log_{10}(\frac{b_2}{b_1}) 
\end{align}
since all other parameters are the same for both bit lengths. Thus, one can compute the SNR$_{\mathrm{dB}}$ value for a varied bit length $b_2$ via the SNR$_{\mathrm{dB}}$ of the bit length $b_1$ by 

\begin{align}
    \mathrm{SNR}_{\mathrm{dB}}(b_2)=\mathrm{SNR}_{\mathrm{dB}}(b_1) + 20 \log_{10}(\frac{b_2}{b_1}).
    \label{equation:fit}
\end{align}
%
%
%
The curve achieved by \cref{equation:fit} with $b_1=10.2\,$nm agrees qualitatively very well with the SNR(bit length) data from Slanovc \textit{et al} \cite{slanovc}. It is thus reasonable to use this equation to include the bit length to the SNR.
\subsection{Combination of damping and bit length \label{section:superiorrole}}
\begin{center}
\begin{table*}
\centering
\begin{tabular}{|>{\centering}m{2.cm}|>{\centering}m{2.cm}|>{\centering}m{3.cm}|>{\centering}m{2.0cm}|>{\centering}m{1.5cm}|c|>{\centering}m{2.0cm}|c|}
\hline
Curie temp. $T_{\mathrm{\textbf{C}}}$ [K] & Damping $\alpha$& Uniaxial anisotropy $k_{u}$ [J/link]& $J_{ij}$ [J/link]& $\mu_{\mathrm{s}}$  [$\mu_{\mathrm{B}}$]  & $v$\,[m/s]&  field duration (fd)\,[ns]&  FWHM\, [nm]\\
    \hline
536.6	 & 0.1&$9.12\times10^{-23}$  & $5.17 \times 10^{-21}$ &1.7&20& 1.0&20\\
    \hline
 \end{tabular}
\caption{Material and write head parameters of a FePt like hard magnetic granular recording medium that were used in former works \cite{areal,fundamental,noisehamr}. }
\label{SNRtableformer}
\end{table*}
\end{center}
\begin{center}
\begin{table*}
\centering
\begin{tabular}{|>{\centering}m{4.0cm}|c|c|c|>{\centering}m{1.5cm}|>{\centering}m{2.0cm}|c|}
\hline
Parameter set & diameter\,[nm]&$T_{\mathrm{peak}}\,$[K] & bit length\, [nm] & $P_{\mathrm{max}}$& $\sigma_{\mathrm{down}}$  [nm] & \,\,SNR\, [dB]\,\,\\
    \hline
ASTC &5&760&10.2 & 0.974 & 0.95 & 17.51\\
	    \hline
Parameters of former works \cite{areal,fundamental,noisehamr}&5&600 &22& 0.984 & 0.384&25.76 \\
	    \hline
	    \hline
ASTC &7&760&10.2 & 0.99 & 0.83 & 15.35\\
	    \hline
Parameters of former works \cite{areal,fundamental,noisehamr}&7&600 &22& 1.0 & 0.44&22.75 \\
	    \hline
 \end{tabular}
\caption{Resulting $P_{\mathrm{max}},\, \sigma_{\mathrm{down}}$ and SNR values for the simulations with ASTC parameters and those used in former simulations. }
\label{SNRwritevariation}
\end{table*}
\end{center}
The simulations with write head and material parameters according to the ASTC are compared to simulations with parameters used in former works \cite{areal,fundamental,noisehamr}. Main differences to the currently used parameters are the bit length, the damping constant, the height of the grain, the exchange interaction, the atomistic spin moment, the full width at half maximum, the head velocity and the field duration. These former parameters are summarized in \Cref{SNRtableformer}. 
Comparing the SNR values of both parameter sets shows that for $d=5\,$nm the SNR is about 8.25\,dB larger for the former used parameters than for the ASTC parameters and for $d=7\,$nm it is $\sim 7.4\,$dB larger. The question is if the damping and bit length variation can fully explain this deviation.\\
Increasing the damping constant from $\alpha=0.02$ to $\alpha=0.1$, yields about $+2.25\,$dB for $d=5\,$nm and $+0.72\,$dB for $d=7\,$nm. Additionally, with the calculations from \Cref{section:bit length}, one can show that by changing the bit length from $b_1=10.2\,$nm to $b_2=22\,$nm gives
\begin{align}
    \mathrm{SNR}_{\mathrm{dB}}(b_2)=\mathrm{SNR}_{\mathrm{dB}}(b_1) + 6.85\,\mathrm{dB}.
\end{align}
Combined, this shows that the difference in the SNR can be attributed entirely to the damping and the bit length enhancement. Moreover, simulations where the other material and write head parameters are changed one by one confirm this findings. The other write head and material parameters that are changed in the simulations have only minor relevance on the SNR compared to the damping constant and the bit length.
\section{Conclusion}
To conclude, we investigated how the damping constant affects the SNR. The damping constant was varied between $\alpha=0.01$ and $\alpha=0.5$ for two different grain sizes $d=5\,$nm and $d=7\,$nm and the SNR was determined. In practice, the damping constant of FePt might be increased by enhancing the Pt concentration \cite{iihama_observation_2013,kim_ultrafast_2011}. Another option would be to use a high/low $T_{\mathrm{c}}$ bilayer structure \cite{suess_breaking_2013} and increase the damping of the soft magnetic layer by doping with transition metals  \cite{doping,ingvarsson_tunable_2004,fassbender_structural_2006,bailey_control_2001,rantschler_effect_2007}.
An interesting finding of the study is the enormous SNR improvement of 6\,dB that can be achieved for $5\,$nm-grains when enhancing the damping constant from $\alpha=0.01$ to $\alpha=0.1$ and beyond. It is reasonable that the SNR improves with larger damping. This results from the oscillatory behavior of the magnetization for small damping during switching. In fact, smaller damping facilitates the first switching but with larger damping it is more likely that the grain will switch stably during the cooling of the thermal pulse \cite{wood2002recording}. This leads to a smaller switching time distribution for larger damping constants and in the further course to higher SNR values. However, an increase of the duration of the heat pulse due to a smaller head velocity or an increase of the field strength can improve the SNR even for smaller damping constant. \\
Furthermore, the results display a SNR saturation for damping constants $\alpha \ge 0.1$. This SNR saturation can be explained with the saturation of the maximum switching probability and the only marginal change of the down-track jitter for $\alpha \ge 0.1$. Indeed, one can check that for shorter pulse widths and smaller field strength, the behavior is different and the SNR does not saturate. In this case, the SNR rises for increasing damping constants. Summarizing, the SNR saturation for a varying damping constant depends strongly on the used field strength and the duration of the heat pulse.\\
The qualitative behavior for 7\,nm-grains is the same. Interestingly, the SNR change for a varying damping constant is not as significant as for grains with $d=5\,$nm. This results from the higher maximum switching probability and the smaller down-track jitter $\sigma_{\mathrm{down}}$ for 7\,nm-grains even for small damping constants. This is as expected since larger grain sizes lead to an elevated maximum switching probability \cite{fundamental} and smaller transition jitter \cite{zhu2013understanding} compared to smaller grain sizes. This limits the possible increase of the recording performance in terms of $P_{\mathrm{max}}$ and $\sigma_{\mathrm{down}}$ and thus the possible SNR gain. Additionally, the SNR saturation value is smaller for 7\,nm-grains since one bit consists of fewer grains.\\
The overall goal was to explain the decrease of the SNR by about $8.25$\,dB and $7.4\,$dB for $d=5\,$nm and $d=7\,$nm, respectively, when changing from recording parameters used in former simulations \cite{areal,fundamental,noisehamr} to the new ASTC parameter. Indeed, together with the bit length variation, the SNR variation could be fully attributed to the damping enhancement. The other changed parameters like the atomistic spin moment, the system height, the exchange interaction and the full width at half maximum have only a minor relevance compared to the influence of the damping $\alpha$ and the bit length.\\
In fact, the variation of the bit length gave the largest SNR change. However, since an increase of the bit length is not realistic in recording devices, the variation of the material parameters, especially the increase of the damping constant, is a more promising way to improve the SNR.

\section{ACKNOWLEDGEMENTS}
The authors would like to thank the Vienna Science and Technology Fund (WWTF) under grant No. MA14-044, the Advanced Storage Technology Consortium (ASTC), and the Austrian Science Fund (FWF) under grant No. I2214-N20 for financial support. The computational results presented have been achieved using the Vienna Scientific Cluster (VSC).


\end{document}